\documentclass[amsmath,amssymb,aps,onecolumn,superscriptaddress,nofootinbib]{revtex4-2}

\usepackage[english]{babel}
\usepackage[utf8]{inputenc}
\usepackage{xr-hyper}

\usepackage[colorlinks]{hyperref}
\usepackage[T1]{fontenc}
\usepackage[dvips]{graphicx}
\usepackage{amsmath}
\usepackage{amsfonts}
\usepackage{color}
\usepackage{ulem}
\usepackage{float}
\usepackage[caption=false, justification=justified]{subfig}
\usepackage{url}
\usepackage{ulem}
\usepackage[top=2cm, bottom=2cm, left=2cm, right=2cm]{geometry}

\makeatletter

\hypersetup{filecolor=blue}
\hypersetup{citecolor=blue}
\hypersetup{urlcolor=blue} 

\hypersetup{linkcolor=blue}

\begin{document}

\title{Quantifying the large contribution from orbital Rashba effect to the effective damping-like torque on magnetization} 


\author{S. Krishnia}
\affiliation{Laboratoire Albert Fert, CNRS, Thales, Université Paris-Saclay, 91767, Palaiseau, France}
\author{B. Bony}
\affiliation{Laboratoire Albert Fert, CNRS, Thales, Université Paris-Saclay, 91767, Palaiseau, France}
\author{E. Rongione}
\affiliation{Laboratoire Albert Fert, CNRS, Thales, Université Paris-Saclay, 91767, Palaiseau, France}
\author{L. Moreno Vicente-Arche}
\affiliation{Laboratoire Albert Fert, CNRS, Thales, Université Paris-Saclay, 91767, Palaiseau, France}
\author{T. Denneulin}
\affiliation{Ernst Ruska-Centre for Microscopy and Spectroscopy with Electrons (ER-C 1) and Peter Grünberg Institut (PGI-5),Forschungszentrum Jülich GmbH, 52425 Jülich, Germany}
\author{Y. Lu}
\affiliation{Ernst Ruska-Centre for Microscopy and Spectroscopy with Electrons (ER-C 1) and Peter Grünberg Institut (PGI-5),Forschungszentrum Jülich GmbH, 52425 Jülich, Germany}
\author{R. E. Dunin-Borkowski}
\affiliation{Ernst Ruska-Centre for Microscopy and Spectroscopy with Electrons (ER-C 1) and Peter Grünberg Institut (PGI-5),Forschungszentrum Jülich GmbH, 52425 Jülich, Germany}
\author{S. Collin}
\affiliation{Laboratoire Albert Fert, CNRS, Thales, Université Paris-Saclay, 91767, Palaiseau, France}
\author{A. Fert}
\affiliation{Laboratoire Albert Fert, CNRS, Thales, Université Paris-Saclay, 91767, Palaiseau, France}
\author{J.-M. George}
\affiliation{Laboratoire Albert Fert, CNRS, Thales, Université Paris-Saclay, 91767, Palaiseau, France}
\author{N. Reyren}
\affiliation{Laboratoire Albert Fert, CNRS, Thales, Université Paris-Saclay, 91767, Palaiseau, France}
\author{V. Cros}
\email{vincent.cros@cnrs-thales.fr}
\affiliation{Laboratoire Albert Fert, CNRS, Thales, Université Paris-Saclay, 91767, Palaiseau, France}
\author{H. Jaffr\`es}
\email{henri.jaffres@cnrs-thales.fr}
\affiliation{Laboratoire Albert Fert, CNRS, Thales, Université Paris-Saclay, 91767, Palaiseau, France}

\date{\today}

\begin{abstract}
{The generation of large spin currents, and the associated spin torques, which are at the heart of modern spintronics, have long been achieved by charge-to-spin conversion mechanisms, i.e. the spin Hall effect and/or the Rashba effect, intrinsically linked to a strong spin-orbit coupling. Recently, a novel path has been predicted and observed for achieving significant current-induced torques originating from light elements, hence possessing a weak spin-orbit interaction. These findings point out to the potential involvement of the orbital counterpart of electrons, namely the orbital Hall and orbital Rashba effects. In this study, we aim at quantifying these orbital-related contributions to the effective torques acting on a thin Co layer in different systems. First, in  Pt|Co|Cu|AlOx stacking, we demonstrate a comparable torque strength coming from the conversion due to the orbital Rashba effect at the Cu|AlOx interface and the one from the effective spin Hall effect in bottom Pt|Co system. Secondly, in order to amplify the orbital-to-spin conversion, we investigate the impact of an intermediate Pt layer in Co|Pt|Cu|CuOx. From the Pt thickness dependence of the effective torques determined by harmonic Hall measurements complemented by spin Hall magneto-resistance and THz spectroscopy experiments, we demonstrate that a large orbital Rashba effect is present at the Cu|CuOx interface, leading to a twofold enhancement of the net torques on Co for the optimal Pt thickness. Our findings not only demonstrate the crucial role that orbital currents can play in low-dimensional systems with weak spin-orbit coupling, but also reveal that they enable more energy efficient manipulation of magnetization in spintronic devices.}
\end{abstract}

\maketitle

\section{Introduction}

The manipulation of magnetization through electrical means at room temperature,
usually achieved through torques originating from pure spin current~\cite{miron2011,liu2012,liu2012bis,manchon2019} and/or non-equilibrium spin density, and involving heavy metals or oxides~\cite{Noel2020,sanchez2013}, holds huge potential for energy-efficient and high-speed spintronic device applications. These encompass spin-orbit torque based magnetic random access memory (SOT-MRAM)~\cite{Kateel2023,Nicholas2021}, logic gates~\cite{Raab2022}, skyrmion-based racetracks~\cite{Fert2013,Fert2017}, as well as emerging neuromorphic computing applications ~\cite{Romera2018,Song2020,Zahedinejad2022,Sethi2023}. However, despite more than one decade of research for current-induced torque optimization, recent observations in systems composed of light elements \textit{i.e.} weak spin-orbit interaction, have challenged the conventional understanding regarding their origin ~\cite{Otani2021,LeeNatCom2021,ElHamdi2023,Krishnia2023}. In fact, the spin current responsible for these torques were previously uniquely associated to the spin Hall effect (SHE) in heavy 5\textit{d} transition metals (TM)~\cite{miron2010,manchon2015} or to the spin Rashba-Edelstein effect at interfaces with broken inversion symmetry~\cite{rashba1983}, two effects for which the relativistic spin-orbit coupling (SOC) plays a crucial role. 
These findings in systems with weak SOC represent a paradigm shift by underscoring the significance of both electronic spin and orbital degrees of freedom at equal footing and even accentuate the generation of orbital currents as a potential avenue for a more efficient magnetization manipulation ~\cite{Dongwook2018,GoEPL}.

\medskip

Notably, the response of orbital angular momentum (OAM) to the application of a charge current has been calculated to exceed that of pure spin angular momentum in certain cases~\cite{Leandro2022}. Very interestingly, unlike spin angular momentum, OAM does not necessarily require SOC, thus expanding the range of materials and interface design possibilities for current-induced magnetization control~\cite{Go2017}. In particular, the direct experimental observations of the OAM accumulation in a light element such as titanium~\cite{Choi2023} and more recently in chromium~\cite{Kawakami2023}, \textit{via} the orbital Hall effect (OHE) highlight its relevance. Furthermore, clear experimental demonstrations of current-induced torques at inversion-asymmetric light metal/ferromagnet interfaces have provided crucial insight into the non-equilibrium OAM phenomenon~\cite{Otani2021,Krishnia2023}. Two predominant mechanisms, OHE~\cite{Dongwook2018} and the orbital Rashba-Edelstein effect (OREE)~\cite{Otani2021}, have been proposed to explain these intriguing observations. On the one hand, OHE entails the flow of chiral OAM in opposing directions, perpendicular to the charge current, thereby generating an orbital current. On the other hand, OREE accounts for the accumulation of chiral OAM at inversion-asymmetric interfaces. Such orbital current and/or accumulated OAM may subsequently diffuse and propagate within an adjacent layer, being a ferromagnet or an heavy material. This feature has been recently evidenced by the longer orbital current decoherence length ~\cite{Sala2022,PRLDing,ding2022} compared to the corresponding spin length. However, a fundamental difference is that, contrary to spin current, the orbital current itself cannot interact directly with the magnetization through exchange mechanism, and has to involve the spin-orbit interactions. In this regard, given the relatively weak SOC of ferromagnetic layers, an intermediate \textit{orbital-to-spin conversion} layer has to be strategically incorporated between the ferromagnetic layer and the orbital current source, \textit{e.g} Pt~\cite{Lee2021,PRLDing} or rare-earth elements with strong SOC such as Gd or Tb~\cite{Lee2021,Sala2022}. Therefore, the overall effectiveness of the torques acting on the magnetic layer hinges on the strength of spin-orbit interaction \textit{i}) in the orbital layer like Cu* in Refs.~\cite{Otani2023,Fert2023}, \textit{ii}) in the heavy metal spacer layer as well as \textit{iii}) in the ferromagnetic layer. Consequently, a systematic study on new materials is required and it becomes an important issue to be tackled from both fundamental and technological point of views.

\medskip

In this study, we report on the experimental observation of the strong enhancement of the damping-like (DL) current-induced torques obtained in sputtered Pt|Co|Cu|AlOx and Co|Pt|CuOx samples with respective out-of-plane and in-plane magnetic anisotropy and compared to reference SHE Co|Pt systems. This enhancement of DL  torque is explained by the strong OREE at the light element and an oxide interface. In that sense, this study differs from our recently published work~\cite{Krishnia2023}, which focused on the huge exaltation of the field-like torque (FLT) component observed in Pt|Co|Al|Pt systems due to OREE occurring at the ferromagnet-light metal (Co|Al) interface. Furthermore, in the Co|Pt|Cu|CuOx series, we investigate the Pt thickness dependence of the subsequent orbital-to-spin conversion and demonstrate an additional impact on the DL torques.
By harnessing such OREE in light transition metal (light-TM) based structures, we achieve a noteworthy twofold enhancement in the effective torques on the Co layer, as validated through harmonic Hall as well as orbital Hall magnetoresistance (OMR) measurements. Our conclusion reveals new aspects of how the orbital properties of conduction electrons may influence the magnetization dynamics, opening up new avenues to control it in future energy-efficient spintronic devices.

\section{Material systems evidencing large orbital Rashba effects}

The different types of multilayers studied in this work have been grown at room temperature using dc magnetron sputtering onto thermally oxidized Si|SiO$_2$ (280~nm) substrates. 
A more detailed description of the growth conditions by sputtering deposition and device fabrication can be found elsewhere~\cite{Krishnia2023}. The sample series having an Al or Cu films as top layer are naturally oxidized in air, thus forming respectively AlOx or Al|AlOx and CuOx or Cu|CuOx interfaces depending on the initial thickness of Al or Cu. In the following, these samples will be marked with a star symbol ($*$). Note that for the samples having Pt (3 nm) as top layer, we know from a previous work using XPS characterization, that the underlying material i.e. Al or Cu, is fully protected from oxidization~\cite{Krishnia2023}. For the electrical and torque measurements, the plain multilayers are designed and fabricated into 5~$\mu$m wide Hall bar structures by using optical lithography and Ar$^+$ ion-milling technique.

\begin{figure}
         \centering
         \includegraphics[width=0.9\textwidth]{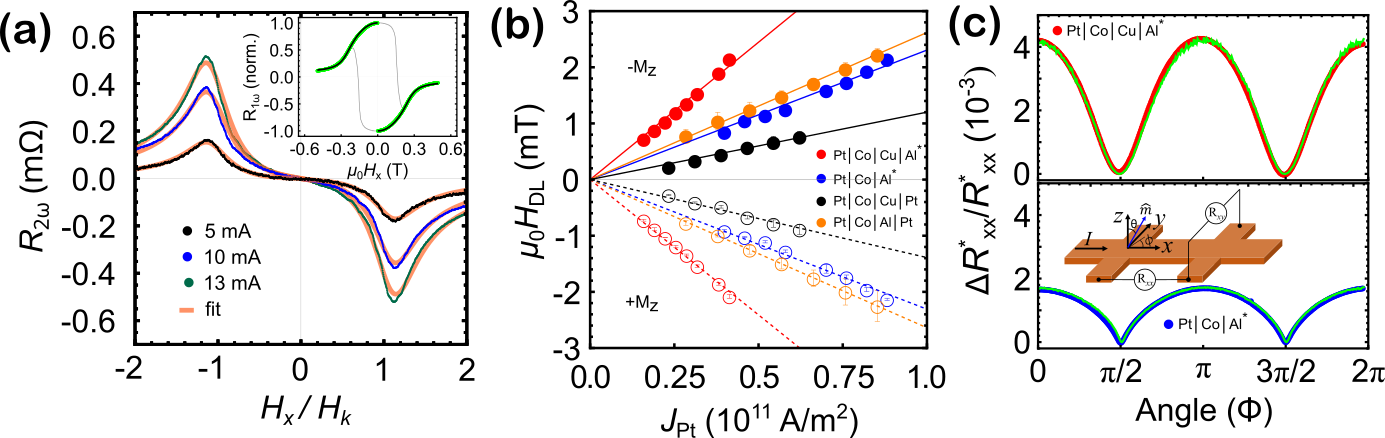}
\caption{(a) Second-harmonic Hall resistance ($R_{2\omega}$) as a function of the in-plane magnetic field ($H_{\rm{x}}$), normalized by the anisotropy field ($H_{\rm{k}}$), in Ta(5)|Pt(8)|Co(1.4)|Cu(5)|Al(2)* for three different currents. The solid orange lines represent the fits using Eq.\ref{eq1}. Inset shows the normalized first-harmonic Hall resistance ($R_{\omega}$) as a function of the in-plane magnetic field and the fit in green using Stoner-Wohlfarth model. (b) Comparison of $H_{\rm{DL}}$ as a function of $J_{\rm{Pt}}$ in Ta(5)|Pt(8)|Co(1.4)|Cu(5)|Al(2)* (red), Ta(5)|Pt(8)|Co(1.4)|Al(2)* (blue), Ta(5)|Pt(8)|Co(1.4)|Al(3)|Pt(3) (orange) and Ta(5)|Pt(8)|Co(1.2)|Cu(1.4)|Pt(3) (black). The $H_{\rm{DL}}$ values in Ta(5)|Pt(8)|Co(1.2)|Cu(1.4)|Pt(3) are multiplied by a factor of 0.85 to normalize them with 1.4~nm Co thickness.  All samples exhibit out-of-plane magnetic anisotropy. The open and filled circles represent two opposite magnetization directions. (c) Spin Hall magneto-resistance measurements in Ta(5)|Pt(8)|Co(1.4)|Cu(5)|Al(2)* (top panel) and Ta(5)|Pt(8)|Co(1.4)|Al(2)* (bottom panel) samples. The change in the longitudinal resistance in the same geometry extracted from the anomalous Hall resistance is shown in green (see text). The schematic of the measurement is also shown.
}
\label{CuAl}
\end{figure}

\subsection{Ta|Pt|Co|Cu|Al* samples (perpendicular magnetic anisotropy)}

We first investigate the current-induced torques in Ta(5)|Pt(8)|Co(1.4)|Cu(5)|Al*(2) deposited on Si|SiO$_2$. The number in parenthesis indicates the thickness in nm. The results are then compared with the ones obtained in Ta|Pt|Co|Cu|Pt, Ta|Pt|Co|Al|Pt and Ta|Pt|Co|Al* samples that serve as references for current-induced torques measurements (see also our recent work by Krishnia \textit{et al.}~\cite{Krishnia2023}.

\subsubsection{Experimental estimation of the damping-like torque (DL torque)}

\medskip

In order to quantify the current-induced torques, we have used the harmonic Hall measurement technique. In practice, an alternating current (ac) of frequency $\omega$ = 727~Hz is injected in the Hall bar that generates orbital and/or spin current and subsequently exert torques on the magnetization with two components: the respective damping-like (DL) and field-like (FL) components. The resultant effective fields induce small quasi-static oscillations of magnetization around its equilibrium position, synchronized  with the ac frequency. This dynamic behaviour gives rise to a voltage (or resistance) signal at twice the frequency of ac current (2$\omega$). In the case of samples with perpendicular magnetic anisotropy (PMA), we simultaneously make the acquisition of first and second harmonics of the Hall resistance while sweeping the in-plane magnetic field along the current direction (damping-like geometry) or transverse to the current direction (field-like geometry). Note that in the present study, we focus on the analysis of the damping-like effective field ($H_{\rm{DL}}$), as we find that it is largely the dominant one for the Pt|Co|Cu|Al* sample under investigation. The in-plane magnetic field dependence of the second harmonic Hall resistance ($R_{2\omega}$) allows to extract the DL component of the torque and given by ~\cite{KrishniaPRA2021}:

\begin{equation}
R_{2\omega}=-\frac{1}{2} \sin(\theta)\left(\frac{H_{\rm{DL}} R_{\rm{AHE}}}{H_{\rm{K}} \cos(2\theta)-H_{\rm{x}} \sin(\delta \theta_{\rm{H}}-\theta)}+\frac{2H_{\rm{FL}}R_{\rm{PHE}}~\sec(\delta \theta_H)\sin(\theta)}{H_{\rm{x}}}\right),
\label{eq1}
\end{equation}
where $H_{\rm{DL}}$ and $H_{\rm{FL}}$ are the respective DL and FL effective fields, $H_{\rm{k}}$ is the anisotropy field, $H_{\rm{x}}$ is the in-plane magnetic field, $\theta$ is the angle of magnetization from out-of-plane axis, $\delta\theta_H$ is the tilt angle of the in-plane magnetic field, and $R_{\rm{AHE}}$ and $R_{\rm{PHE}}$ are the anomalous and planar Hall resistances. $R_{\rm{AHE}}$, $\theta$, $H_{\rm{k}}$ and $\delta\theta_{\rm{H}}$ are extracted by fitting the 1$^{st}$ harmonic Hall resistance using the Stoner-Wohlfarth model as shown in the inset of Fig.\ref{CuAl}b and as explained in our previous work~\cite{KrishniaPRA2021,Krishnia2023}.

\medskip

In Fig.~\ref{CuAl}(a), we show the variation of $R_{2\omega}$ as a function of the in-plane magnetic field in the DL geometry for various currents in the Pt(8)|Co(1.4)|Cu(5)|Al(2)* sample along with the fits using Eq.\ref{eq1}. The extracted $H_{\rm{DL}}$ \textit{vs.} the current density injected in the 8~nm bottom Pt ($J_{\rm{Pt}}$) are shown in  Fig.~\ref{CuAl}(b) (red circles). Such local current density has been evaluated through the measurement of both longitudinal $R_{\rm{xx}}$ and transverse $R_{\rm{xy}}$ (or  $R_{\rm{AHE}}$) within a parallel resistor scheme as described hereafter in more details. The $H_{\rm{DL}}$ exhibits a linear increase with $J_{Pt}$ and undergoes a sign change relative to the magnetization direction in agreement with symmetry arguments. The interesting outcome lies in the significant variation of $H_{\rm{DL}}$ magnitude depending on the specific top light-metal interfaces, that are Pt|Co|Cu|Pt, Pt|Co|Al|Pt and Pt|Co|Al*. First, we find a twofold decrease in $H_{DL}$ in Pt|Co|Cu|Pt (black circles) compared to the two reference samples, \textit{i.e.} Pt|Co|Al|Pt (orange circles) and Pt|Co|Al* (blue circles). This reduction can be explained, to a large extend, by an opposite SHE contribution from the top 3~nm Pt given the significant electronic and spin transmission across Cu. In our previous work, such a partial compensation in the SHE between bottom and top Pt has been very precisely evaluated to be 60\% for a characteristic spin-diffusion length $\lambda_{sf}^{\rm{Pt}}$ in Pt of 1.5~nm~\cite{Krishnia2023}; and 50\% for $\lambda_{sf}^{\rm{Pt}}$ in Pt of 1.75~nm, as observed here. This result also highlights that a metallic Cu layer, free of any oxide interface, cannot directly contribute significantly to any OHE but rather serves as a current shunting pathway~\cite{Krishnia2023}. 

\medskip

A second remarkable result observed in  Fig.~\ref{CuAl}(b) is the twofold increase of $H_{\rm{DL}}$ in Pt|Co|Cu|Al* (red circles) compared to the reference samples, \textit{i.e.} Pt|Co|Al|Pt (orange circles) and Pt|Co|Al* (blue circles), which cannot be explained by only considering a contribution from SHE. In this case, it is worth mentioning that the bottom Pt|Co interface remains identical in all these samples and thus, no significant change in the SHE action originating from the bottom Pt layer is expected. Further, we note that Cu and Al are light metals with negligible SOI, so they are not expected to generate a sizeable pure spin current through SHE. However, as discussed previously, recent experimental findings indicate the potential generation of an orbital current at naturally oxidized Cu|CuOx interface~\cite{PRLDing}, largely contributing to the overall net torques. Moreover, we have recently reported on SOTs properties of Pt(t$_1$)|Co|Pt(t$_2$), Pt|Co|Al|Pt and Pt|Co|Cu|Pt series. We clearly demonstrated that Pt(8)|Co(1.4)|Al(3)|Pt(3) involving thick Co (1.4~nm) and Al (3~nm) layers can be regarded as a 'reference' SHE sample in which the $H_{\rm{DL}}$ predominantly originates from the SHE in the bottom Pt as played by the simpler Pt|Co bilayer~\cite{Krishnia2023}. We may then associate, at this stage, the twofold rise in  $H_{\rm{DL}}$ in Pt|Co|Cu|Al* compared to Pt|Co|Al* or/and Pt|Co|Al|Pt to an orbital current generation at the Cu|Al* interface. 

\subsubsection{Longitudinal magnetoresistance: SMR and/or OMR}

In addition to the harmonic Hall measurements, we further validate the observed increase in $H_{DL}$ in Pt|Co|Cu|Al* through magnetoresistance measurements (MR). To this aim, we measured MR in the so-called spin-Hall magnetoresistance geometry (SMR)~\cite{HayashiSMR,PRBChen,Husain2023} with the current being injected along the $\hat\rm{{x}}$ direction and $\hat{z}$ is the normal direction to the layers. In this case, the magnetoresistive change $\Delta R_{\rm{xx}}$ as a function of the angle between the magnetization and the out-of-equilibrium spin or orbital angular momentum polarization (OAM) direction (generated along the in-plane $\hat{y}$ direction), reflects how the angular-momentum current is reflected and/or transmitted at a specific magnetic/non-magnetic interface. Any additional ('\textit{add}') contribution originating from the orbital current generated at the Cu|Al* interface, later on named OMR, shall result in a change in the SMR according to $\frac{\Delta R_{\rm{xx}}}{R_{\rm{xx}}}=\frac{\Delta R_{\rm{xx}}^{\rm{SMR}}}{R_{\rm{xx}}^{\rm{Pt|Co}}}+\frac{\Delta R_{\rm{xx}}^{\text{add.}}}{R_{\text{add}}^{\rm{Co|top}}}$ ('\textit{top}' means here Al*, Cu|Al*, Cu|Pt or Al|Pt). We simultaneously measure both the longitudinal ($R_{\rm{xx}}$) and transverse ($R_{\rm{xy}}$) (AHE) resistances by rotating the sample in the $(\hat{z}-\hat{y})$ plane under the application of a fixed magnetic field of 670~mT, larger than the anisotropy field (see Fig.\ref{CuAl}(c)). In this geometry, the magnetization consistently remains perpendicular to the direction of the charge current flow, thereby eliminating any contribution from  anisotropic magnetoresistance (AMR). Consequently, any variation in $\Delta R_{\rm{xx}}$ solely arises from SMR-like effect as $\Delta R_{\rm{xx}}=\Delta R_{\rm{SMR}}(1-m_y^2)$. 

\medskip

In Fig.\ref{CuAl}(c), we show in red the SMR ratio, $\Delta R_{\rm{xx}}/R_{\rm{xx}}$ \textit{vs.} the rotation angle for Pt|Co|Cu|Al* (red in top panel) and for Pt|Co|Al* (blue in bottom panel) according to $\Delta R_{\rm{xx}} = R_{\rm{xx}} (\hat{m}\parallel \hat{z} ~or~  \hat{m}\perp\hat{\sigma})$ - $R_{\rm{xx}} (\hat{m}\parallel \hat{y} ~or~  \hat{m}\parallel\hat{\sigma})$ with $\hat{m}$ and $\hat{\sigma}$ being respectively the unit vector along the magnetization and spin polarization directions, respectively. For comparison, we also include in Fig.\ref{CuAl}(c) the effective SMR from the $R_{\rm{AHE}}$ data (in green) according to a relationship $R_{\rm{xx}}=R_{\rm{xx}}^0+R_{\rm{SMR}}\left(1-\left(\frac{R_{\rm{xy}}}{R_{\rm{AHE}}}\right)^2\right)$, giving an excellent quantitative agreement between the two angular variations. The increase of the SMR ratio by a factor of more than two between the two samples witnesses the existence of an OAM polarization originating from the Cu|Al* interface. Such orbital accumulation is then after converted either into a charge current in the Co layer and/or into an intermediate spin-current at the bottom Pt|Co interface by SOC (orbital-to-spin filtering effect in Pt) before being released into a charge current \textit{via} inverse spin-Hall effect (ISHE). 

\subsection{Co|Pt|Cu* samples (In-plane magnetized samples)}

In this section, we focus on the study of the net DL torques acting on the Co magnetization in Co(2)|Pt($t_{Pt}$)|Cu*(3) samples exhibiting an in-plane magnetic anisotropy. The field required to saturate the magnetization along the out-of-plane direction is about 950~mT (see SI-III). Note that for sample series, we deliberately omit any Ta|Pt buffer layer and directly deposit the multilayers on Si|SiO$_2$ for avoiding spurious spin-current from any bottom SHE Pt layer, strongly competing with the orbital component. In order to determine the actual impact of orbital currents on the effective torque, we have compared the results with the ones resulting from various non-magnetic metals, including Pt|Cu*, Cu*, Pt and Pt|Al*. The last two cases are considered as reference samples. These capping layers serve two primary purposes: first to prevent oxidation of the Co layer and second to act as a source of spin and/or orbital current. 

\begin{figure}
         \centering
         \includegraphics[width=0.8\textwidth]{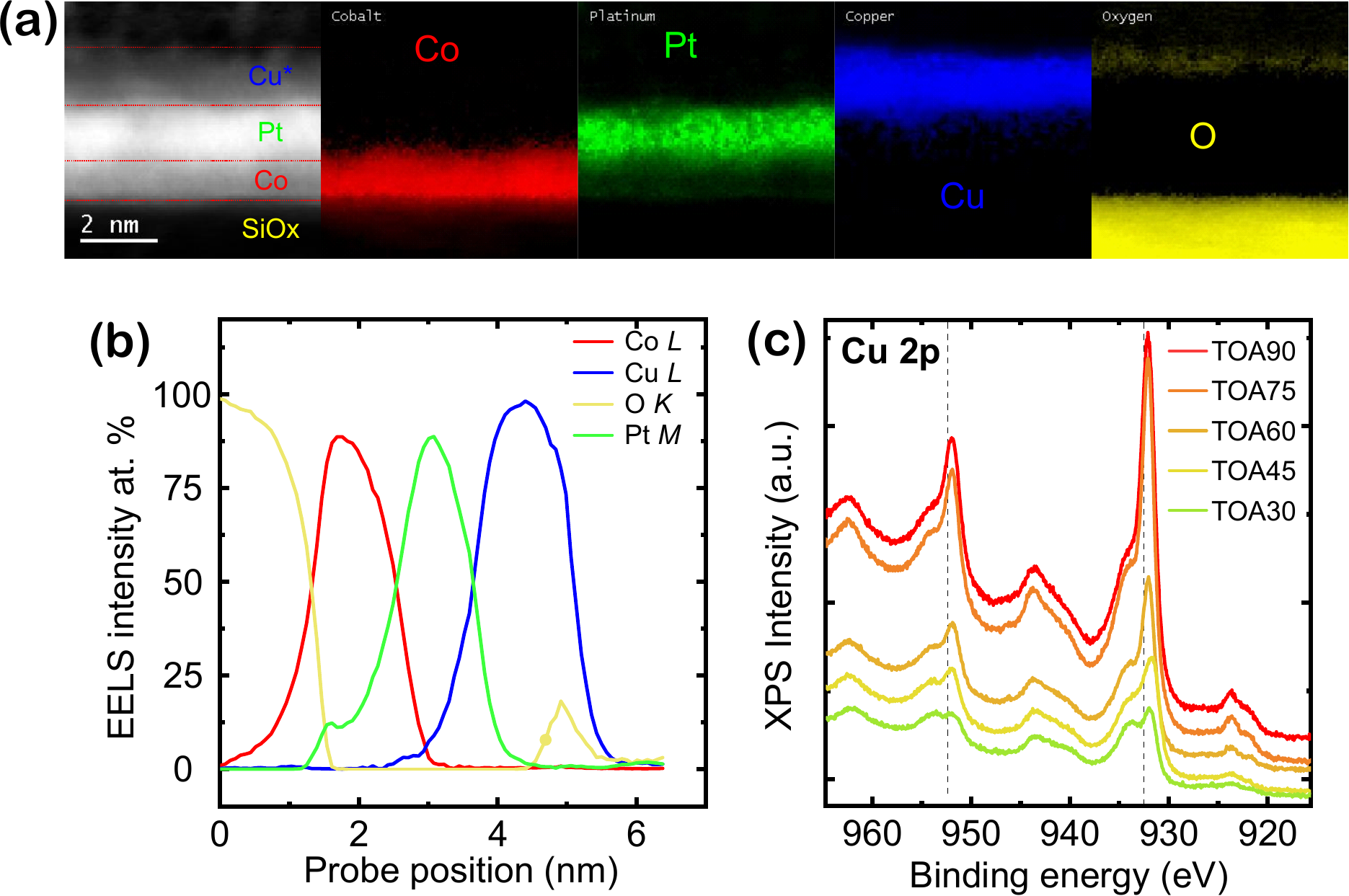}
\caption{(a)HAADF STEM image of Co(2)|Pt(3)|Cu(3)* sample grown in SiOx. The EELS element mapping displaying the spatial elemental distribution of Co (red), Pt (green), Cu (blue) and Oxygen (yellow) is alsow shown. (b) Profiles of atomic fraction of Co (red), Pt (green), Cu (blue) and Oxygen (yellow) measured along the growth direction and averaged along the direction parallel to the interfaces. 
(c) Take-off angle (TOA) dependence of XPS spectra in Co(2)|Pt(2)|Cu(3)* for Cu 2\textit{p} transition. The probing depth increases with TOA being maximum for TOA90. 
}
\label{TEM_XPS}
\end{figure}

\subsubsection{Structural and chemical characterization}

Concerning this particular series, in which the 2~nm thin Co layer is directly grown on Si|SiO$_2$, it is essential to get some insights about the structural and chemical properties of the samples in order to corroborate the conclusions revealed from torque behavior. In Fig.~\ref{TEM_XPS}(a), we show a cross-section image of the Co(2)|Pt(2)|Cu*(3) sample obtained using high-angle annular dark-field scanning transmission electron microscopy (HAADF STEM). 
The spatial distribution along the thickness for each element \textit{i.e.} Co (red), Pt (green), Cu (blue) and Oxygen (yellow) obtained using electron energy loss spectroscopy (EELS) technique is plotted in Fig~\ref{TEM_XPS}(b). These EELS data clearly show first, that each layer is continuous, flat and homogeneous and, second, that only the top part of the Cu layer is oxidized, clearly revealing a pristine Cu of about 1.5~nm thick.

\medskip

In addition, we have also performed some X-ray photo-electron spectroscopy (XPS) at various take-off angles (TOA) in order to gain insight into the distribution of species within the top Cu* layer in Co|Pt|Cu* and Co|Cu* stacks. This TOA corresponds to the angle between the sample surface and the collection axis of the spectrometer : TOA90 is equivalent to normal emission (maximum probing depth), TOA30 is equivalent to a geometry with a higher degree of inclination (minimum probing depth). In case of a gradient of composition in the Cu* layer, the contribution of near-surface species to the spectrum shall be the largest for the geometry with the shallower probing depth (\textit{i.e.}, TOA30). The evolution of the shape of the XPS spectra at different TOA shown in Fig.~\ref{TEM_XPS}(c), notably close to L$_3$ edge (around 932 eV), bring some evidence that a spatial variation of the oxidation state is present within the Cu(3)* top layer with a typical distribution being Cu|Cu$_2$O|Cu(OH)$_2$ along the z-direction. A much more detailed analysis of these XPS characterization is presented in the Supplementary Information (SI-1).

\begin{figure}
         \centering
         \includegraphics[width=0.8\textwidth]{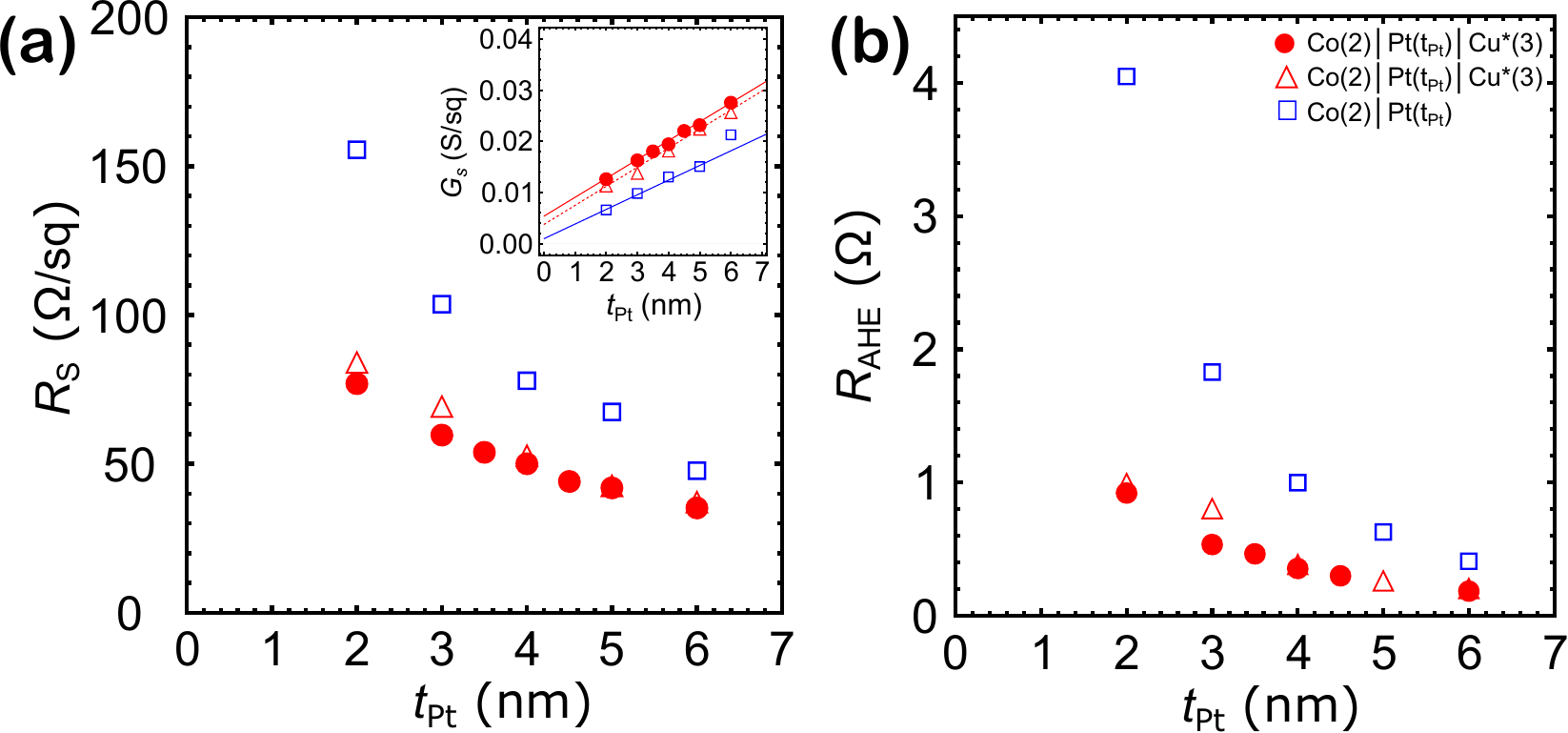}
\caption{ (a) $R_s$ and (b) $R_{AHE}$ as a function of $t_{Pt}$ in Co(2)|Pt($t_{Pt}$) (blue open squares) and in two Co(2)|Pt($t_{Pt}$)|Cu*(3) series of samples (red solid circles and open triangles). The sheet conductance ($G_s=R_s^{-1}$) is shown in the inset of (a).
}
\label{RAHEandRs}
\end{figure}

\subsubsection{Determination of the metallic Cu thickness}

Complementary to the EELS and XPS results presented before, the resistance-area product 
$R_{\rm{AHE}}$ can also be used to check the presence of a thin metallic Cu layer through the estimation of the shunting current as described within a simple \textit{parallel resistor} scheme. The evolution of the sheet resistance $R_s$ as a function of Pt thickness $t_{\rm{Pt}}$ is shown is Fig.~\ref{RAHEandRs}(a) for the two types of samples \textit{i.e.} Co(2)|Pt($t_{\rm{Pt}}$)|Cu(3)* (2 series) and Co(2)|Pt($t_{\rm Pt}$). The slope of the $R_s^{-1}=G_s=G_{\rm{Pt}}+G_{\rm{rest}}=\sigma_{\rm{Pt}} t_{\rm{Pt}}+G_{\rm{rest}}$ \textit{vs} $t_{\rm{Pt}}$ in the inset of Fig.~\ref{RAHEandRs}(a) gives the Pt resistivity $27\pm 2~\mu~\Omega$~cm. The shift of the intercept of the linear variation between the two series allows to extract the product of the Cu*(3) material conductivity times its thickness. 
From this, we deduce a thickness for the pristine metallic Cu layer to be about 1~nm, considering a typical resistivity $\rho_{\rm{Cu}} = 10~\mu \Omega$~cm at room temperature. As a corollary to $R_s$, $R_{\rm{AHE}}$ may also provide additional information about the current shunt mechanism in a same \textit{parallel resistance} scheme. Indeed, as revealed in Fig.~\ref{RAHEandRs}(b), the AHE amplitude severely drops as $t_{\rm{Pt}}$ increases over its typical mean-free path ($\lambda^{\rm{Pt}}$), mainly due to the passive action introduced by the Pt and Cu* layers. From Ref.~\cite{Dang2020}, the standard law of the AHE variation taking into account such current shunting effects reads $R_{\rm{AHE}}^{\rm{Cu}*}(\zeta)=\zeta^2~R_{\rm{AHE}}^{\text{Ref}}$, with $\zeta=\frac{R(\rm{Cu}*)}{R^{\text{(Ref)}}}$ being the ratio of the respective resistances with Cu*/and without the additional shunt~\cite{Dang2020}. This ratio corresponds to the proportion of the CIP-current in the reference AHE system within the whole structure. Such a model aligns well with our experimental results.  For instance, in the case where a 1~nm Cu layer is added to a 2~nm thick Pt current shunt, $R_{s}$ decreases by a factor of 2, while the $R_{\rm{AHE}}=R_{\rm{xy}}$ drops by a factor of 4. Similarly, for $t_{\rm{Pt}}=4$~nm, the drop in $R_{s}$ from $\approx 78~\Omega/\square$ to $\approx 50~\Omega/\square$ (the $\zeta = 1.56$ with corresponding $\zeta^2\approx 2.45$) with Cu* inserted leads to a drop of $R_{\rm{AHE}}$ by $\approx$~2.5 from 1 to 0.4 $\Omega$.

\begin{figure}
         \centering
         \includegraphics[width=0.8\textwidth]{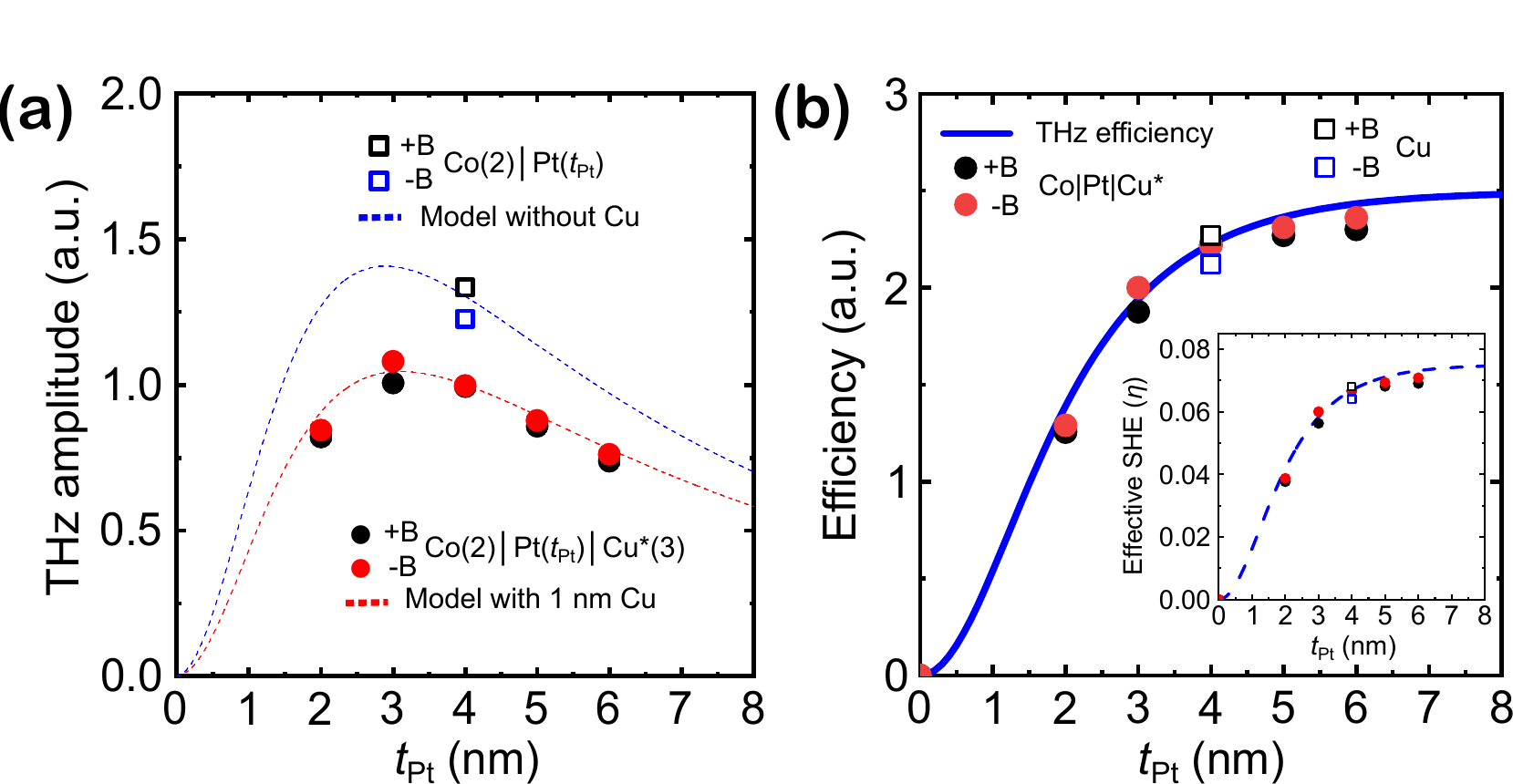}
\caption{ (a) Amplitude of the $E_{\rm{THz}}$  emitted from Co(2)|Pt(t$_{Pt}$)|Cu(3)* (red and black circles) and Co(2)|Pt(t$_{Pt}$) (blue and black open squares) as a function of the Pt thickness for an in-plane saturating plane magnetic field $B\simeq 0.1$~T. The respective lines are the fit corresponding to the product of $\eta \times \mathcal{A}$ (see text). They have been obtained by considering a conductivity of Pt of 4$\times 10^6$ S/m (resistivity $\rho_{\rm{Pt}}=25~\mu\Omega$.cm) and a 1~nm conductive Cu layer of conductivity 10$^7$ S/m (resistivity $\rho_{\rm{Cu}}=10~\mu\Omega$.cm) in Co(2)|Pt(t$_{Pt}$)|Cu(3)*. (b) Plot of the $\eta$ function (inset) exhibiting the efficiency of the spin injection and subsequent THz emission. {\textit{Error bars are given by the size of the dots in the plot}}.
}
\label{THzTDS}
\end{figure}

\medskip

The THz spectroscopy experiments (THz-TDS) in the so-called emission mode is an alternate method to probe the partial metallic character of Cu* within the multilayer. Details of the respective set-up and protocol are provided in the Supplementary Information (SI-II) and in Refs.~\cite{Dang2020,Hawecker2021,Hawecker2022}. 
The THz amplitudes obtained \textit{vs.} the Pt thickness are presented in Fig.~\ref{THzTDS}(a), where the black and red points correspond to the absolute value of E$_{\text{THz}}$ for the two opposite saturating magnetic field orientated in the sample plane and corresponding respectively to the +/- THz waveform polarities. The observed shape of the E$_{\rm{THz}}$ \textit{vs.} $t_{\rm{Pt}}$ plot results from the product of two functions respectively $\eta \times \mathcal{A}$ involving the SCC efficiency $\eta$ and $\mathcal{A}$ describing the effect of both the THz and NIR optical absorptions. 

\medskip

The spin-charge conversion efficiency $\eta$ is a growing function of $t_{\rm{Pt}}$, $\eta(t_{\rm{Pt}})=\theta_{\rm{SHE}}\frac{g_{\uparrow \downarrow} r_s ~\tanh\left(\frac{t_{\rm{Pt}}}{2\lambda_{sf}^{\rm{Pt}}}\right)}{1+g_{\uparrow \downarrow} r_s \coth\left(\frac{t_{\rm{Pt}}}{\lambda_{sf}^{\rm{Pt}}}\right)}$ passing through the origin ($\eta=0$ for $t_{\rm{Pt}}=0$), as displayed in Fig.~\ref{THzTDS}(b) (in THz arbitrary unit) and its inset giving the corresponding dependence of the effective spin-Hall angle $\eta=\theta_{\rm{SHE}}^{\text{eff}}$ \textit{vs.} $t_{\rm{Pt}}$. $\lambda_{sf}^{\rm{Pt}}$ is the hot electron spin relaxation length in Pt, the typical lengthscale over which the ISHE occurs; $r_s=\rho_{\rm{Pt}}\times\lambda_{sf}^{\rm{Pt}}$ is the corresponding spin resistance of Pt~\cite{Dang2020} (we will further call $\tilde{r}_s=g_{\uparrow \downarrow}r_s$, the reduced spin resistance times the spin-mixing conductance). 

\medskip

Accessing the Pt conductivity responsible for the optical NIR absorption allows us to determine both the thickness and the resistivity of the remaining metallic Cu layer. The estimated values are approximately 1~nm and 10~$\mu \Omega$.cm. We have checked that the THz signal emitted from a reference sample that do not contain any Cu and/or Cu|Cu* interface (blue and black points on Fig.~\ref{THzTDS}(b) for the two different field polarities), is also well recovered by the same simulation protocol. This ensures the consistency of both measurements and analyses by fixing the hot electron spin relaxation length $\lambda_{sf}^{Pt}=1.45\pm 0.05$~nm and $\tilde{r}_s =\left( g_{\uparrow \downarrow}r_s\right)=2$. 

\begin{figure}
         \centering
         \includegraphics[width=0.9\textwidth]{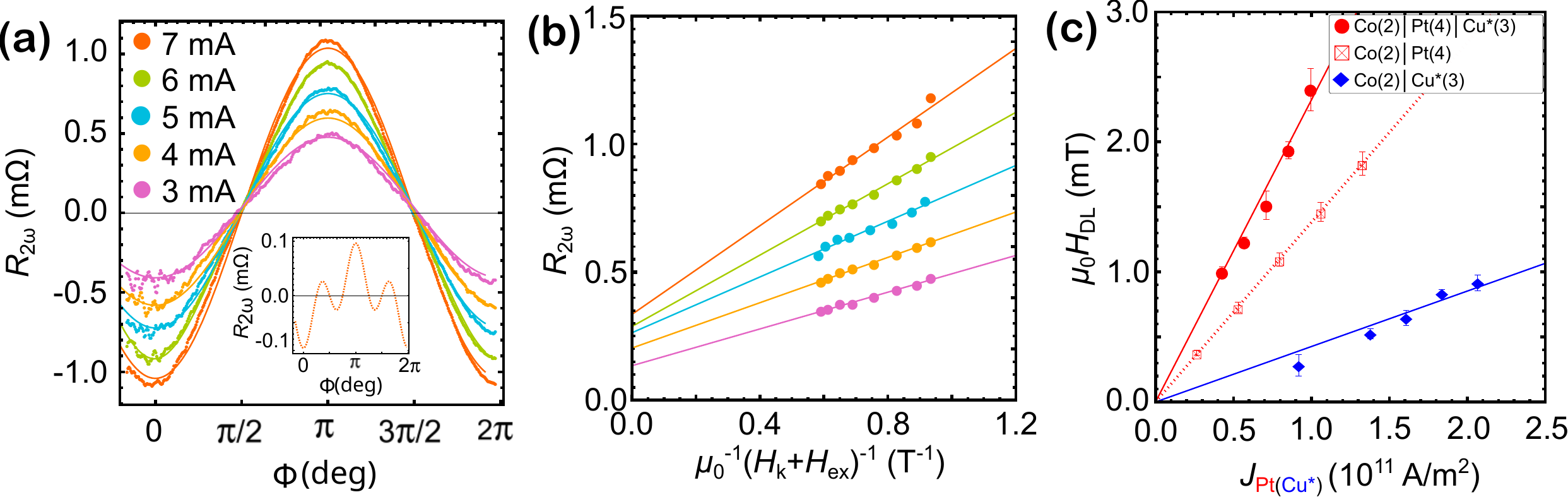}
\caption{(a) $R_{2\omega}$ as a function of in-plane angle between injected current and magnetization direction for various currents in Co(2)|Pt(4)|Cu*(3) sample. The dotted lines are the cosine fits to separate the H$_{FL}$ and H$_{Oe}$ components from the H$_{DL}$ and thermal effects. The residual of cosine fit is shown in the inset.  (b) Amplitude of the cosine fit as a function of effective field for different currents. The corresponding linear fits are shown by linearly fitted lines. (c) The H$_{DL}$ as a function of $J_{Pt}$ in Co(2)|Pt(4) (open squares) and Co(2)|Pt(4)|Cu*(3) samples (solid red  circle).The H$_{DL}$ values in Co(2)|Cu*(3) sample as a function of current density in Cu* are shown by blue solid circles. {\textit{Error bars are estimated by the circle size reported in the plots.}}
}
\label{torques}
\end{figure}

\subsubsection{Quantification of the net DL torques}

To quantify the current-induced torques in samples with in-plane magnetic anisotropy, we have measured $R_{1\omega}$ and $R_{2\omega}$ as a function of the angle ($\phi$) between the current and magnetization by rotating the sample in the plane, under a constant in-plane magnetic field (H$_{\rm{ex}}$). The angular dependence of $R_{2\omega}$ obtained for the Co(2)|Pt(4)|Cu*(3) sample for various currents and a fixed H$_{\rm ex}$ = 206 mT are shown in Fig.~\ref{torques}(a). The variation in $R_{2\omega}$ with respect to $\phi$ arises from a combination of current-induced effective fields, including  $H_{\rm{DL}}$, $H_{\rm{FL}}$ and Oersted ($H_{\rm Oe}$) effective fields, as well as thermal effects. The angular dependence of $R_{2\omega}$ can be expressed as:

\begin{equation}
     R_{2\omega}=\left(R_{\rm AHE}\times \frac{H_{\rm DL}}{H_{\rm k}+H_{\rm ex}}+I\alpha \Delta T\right) \cos(\phi)+\\2R_{\rm PHE}\times(2\cos^3(\phi)-\cos(\phi))\times\frac{H_{\rm FL}+H_{\rm Oe}}{H_{\rm ex}  },
     \label{eq2}
\end{equation}

\noindent where $\alpha$ is the thermal coefficient. In order to separate the two contribution (\textit{resp.} $H_{\rm{DL}}$ + $I \alpha \Delta T$ and $H_{\rm{FL}}$+$H_{Oe}$) appearing in the right hand side of Eq.[\ref{eq2}], we have fitted R$_{2\omega}$ \textit{vs.} $\phi$ curves with cos($\phi$) by forcing it to cross through the points where the second term vanishes and the R$_{2\omega}$ signal arises solely from H$_{\rm{DL}}$ plus the thermal effects. The fittings are shown by dotted lines in Fig.~\ref{torques}(a) for several injected currents and corresponding current densities. The residual of the cosine fit simply gives us the H$_{\rm{FL}}$+H$_{Oe}$ contribution as shown in the inset of Fig.~\ref{torques}(a) obtained for a current I = 5 mA. Since the amplitude of H$_{\rm{FL}}$ (term in $2\cos\left(\phi\right)^3-\cos\left(\phi\right)$ in the right-hand side of Eq.~\ref{eq2}) is one order of magnitude smaller than H$_{\rm{DL}}$ in all our samples (see inset of Fig.~\ref{torques}a), our primary focus in the subsequent analysis will be on the H$_{\rm{DL}}$ torque component. In order to disentangle the contribution from thermal effects, we have also measured R$_{2\omega} $\textit{vs.} $\phi$ scans at several external magnetic fields. The slope of the linear fit (dotted lines) yields the magnitude of H$_{\rm{DL}}\times$ R$_{\rm{AHE}}$, while the intersection corresponds to the thermal contribution ($I\alpha\Delta T$). 

\medskip

In Fig.\ref{torques}(c), we compare $H_{\rm{DL}}$ measured in Co|Cu*, Co|Pt|Cu* and Co|Pt structures as a function of the current density in Pt and in Cu* for samples free of any Pt. First, we find as expected that $H_{\rm{DL}}$ increases linearly with the current density for all the samples, which confirms the reliability of our measurements. Second, we measure a sizable DL torque on the Co layer in a system without any heavy metal interface i.e. Co|Cu* sample (solid blue circles). The amplitude of the damping-like effective field is found to be $\mu_0 H_{\rm{DL}}/J_{\rm{Cu}}$ $\approx$ 0.4~mT/$10^{11}$A/m$^2$ current density in the remaining 1~nm metallic Cu. The existence of such DL torque in Co|Cu* sample hints on the orbital current generation \textit{via} the orbital Rashba-Edelstein effect at the Cu|CuOx interface that, in turn, propagates into Co before being converted into a spin current \textit{via} the SOC of Co. This observation agrees with results from previous studies in NiFe|Cu* bilayer ~\cite{YukioPRL,ding2022}, CoFeB|Cu*~\cite{huang2023}, YIG|Pt|Cu*~\cite {vanwees2024} or in Cu|AlOx~\cite{Otani2021}, hence also emphasizing on the absence of any spin-Rashba interactions in these systems. In the same spirit, more recent results demonstrate the orbital transport and orbital conversion at Cu|Oxide (Cu|MgO) interface by OREE following ultrashort optical excitation~\cite{xu2023}. A third important result in Fig.\ref{torques}(c) that can be emphasized is the observed twofold increase in the $H_{\rm{DL}}$ amplitude upon capping the Co|Pt (red square) bilayer with Cu* (red circle). We find that $H_{\rm{DL}}$ amplitude increases from $\approx$ 1.38~mT/$(10^{11}$~A/m$^2$ in  Co(2)|Pt(4) bilayer structure to $\approx$ 2.31 mT/$(10^{11}$ A/$m^2$) in Co(2)|Pt(4)|Cu*(3) trilayer for $10^{11}$~A/m$^2$ current density in Pt. Thanks to THz-TDS experiments (Fig.~\ref{THzTDS}), we have also checked that such results cannot originate \textit{e.g.} from a change in the effective conductivity of Pt on adding Cu* on top (compared to Co|Pt|air) which would have for effect to enhance the SHE current.

\section{Orbital-to-spin conversion in Pt}

\subsection{Amplification of the orbital Rashba effect through insertion of a thin \textit{via} Pt layer}

\begin{figure}
         \centering
         \includegraphics[width=0.8\textwidth]{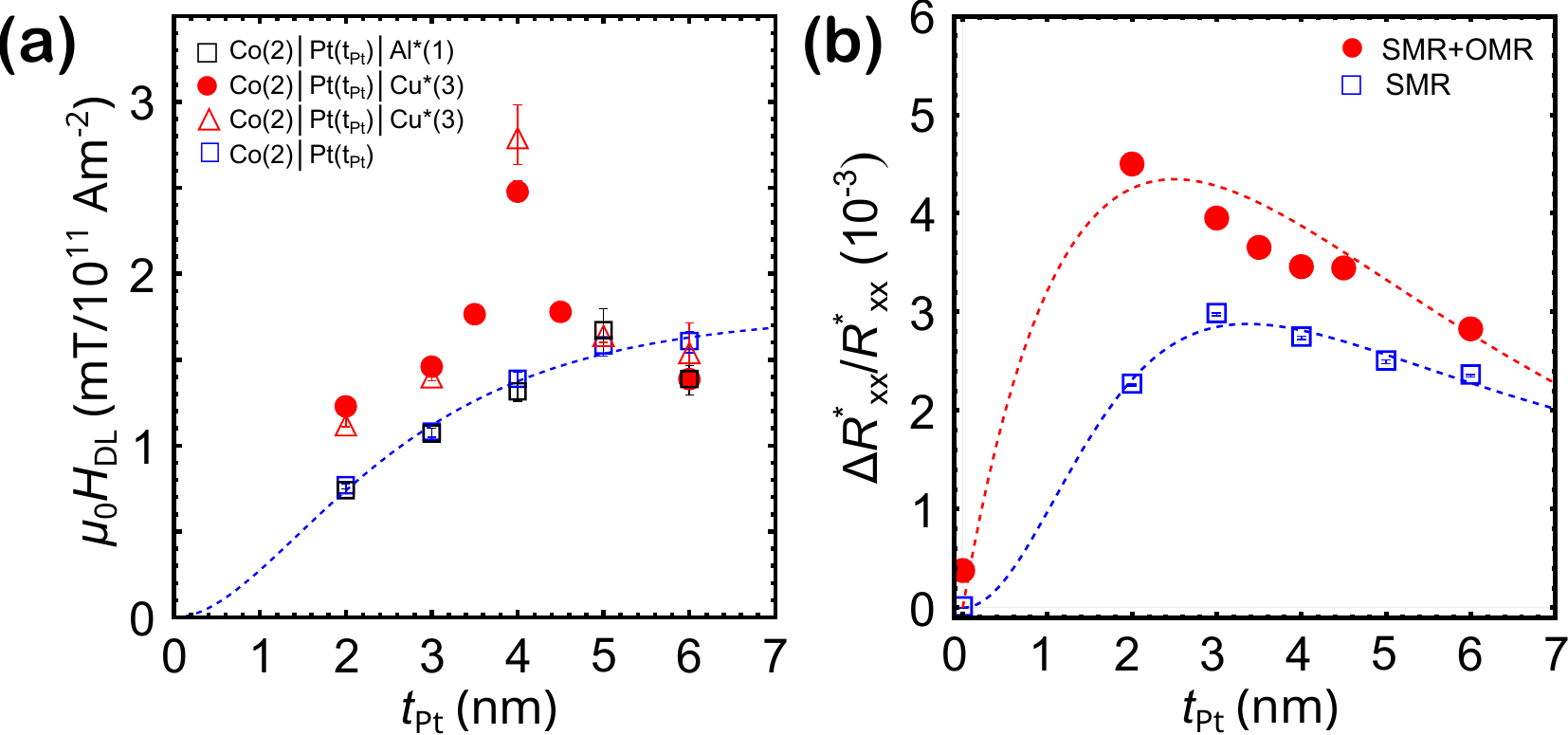}
\caption{ (a) $H_{\rm DL}$ effective fields  normalized by current density in Pt as a function of Pt thickness in Co(2)|Pt($t_{\rm Pt}$) (blue open square), Co(2)|Pt($t_{\rm Pt}$)|AlO$_x$(1) (black open square),  Co(2)|Pt($t_{\rm Pt}$)|Cu(3)* (red circles and open triangles). The blue dotted line is the fit corresponding to the function $\zeta$ defined in the text. (b) Spin Hall magnetoresistance normalized by the longitudinal resistance and considering the current shunt effects in Co(2)|Pt($t_{\rm Pt}$)|Cu*(3) (red circles) and Co(2)|Pt($t_{\rm Pt}$) (open blue square), together with the fits in blue dotted line. The red line is a guide to the eyes calculated from our spin-orbit assisted diffusion model based on Ref.~\cite{kokado2012}}
\label{filtering}
\end{figure}

In order to gain deeper insights about the origin of the strong enhancement of $H_{\rm{DL}}$ in Cu*-based samples, we have performed a systematic investigation of the evolution of $\mu_0 H_{\rm{DL}}$/$J_{\rm{Pt}}$ as a function of the $t_{Pt}$ in the Co(2)|Pt($t_{\rm{Pt}}$)|Cu(3)* series and compare it to the results in the reference series Co(2)|Pt($t_{\rm{Pt}}$). As shown in Fig.~\ref{filtering}(a), the evolution of the normalized DL torque $\mu_0 H_{\rm{DL}}$/$J_{\rm{Pt}}$ for Co(2)|Pt($t_{\rm{Pt}}$) exhibits a steady increase with $t_{\rm{Pt}}$ that eventually starts to saturate at $t_{\rm{Pt}}$=5~nm (blue open square). In Fig.~\ref{filtering}(a) (black open squares), we also include the evolution of $\mu_0 H_{\rm{DL}}$/$J_{\rm{Pt}}$ with $t_{\rm{Pt}}$ recorded for Co(2)|Pt($t_{\rm{Pt}}$)|Al(1)* series. Note that this evolution with $t_{\rm{Pt}}$ followed the same trend as the one obtained from ISHE data obtained from THz-TDS presented in Fig.~\ref{THzTDS}, in which the SHE efficiency \textit{vs.} $\eta(t_{\rm{Pt}})$ follows the function $\eta(t_{\rm{Pt}})=\theta_{\rm{SHE}}\frac{\left(g_{\uparrow \downarrow}r_s\right)\tanh\left(\frac{t_{\rm{Pt}}}{2\lambda_{sf}^{\rm{Pt}}}\right)}{1+\left(g_{\uparrow \downarrow} r_s\right) \coth\left(\frac{t_{\rm{Pt}}}{\lambda_{sf}^{\rm{Pt}}}\right)}$ as previously defined. Note that we have obtained similar evolution in $\eta$ function as a function of $t_{\rm{Pt}}$ in our previous work on thinner ferromagnetic layer with out-of-plane magnetic anisotropy~\cite{Krishnia2023}. From this, we can conclude that the Pt|Al* interface does not play a significant role in contributing to the additional source of DL torques. By fitting $\mu_0 H_{\rm{DL}}$/$J_{\rm{Pt}}$ vs $t_{\rm{Pt}}$ plot (dotted line), we are able to determine the spin diffusion length of Pt, $\lambda_{\text{sf}}^{\text{Pt}}\approx 1.75\pm 0.05$~nm, slightly larger than the previous value (1.45~nm) as found for optically excited carriers in THz-TDS and the bulk spin Hall angle of Pt, $\theta_{\text{SHE}}^{\text{Pt}}=0.21\pm 0.02$. The obtained parameters ($\lambda_{\text{sf}}^{\text{Pt}}$ and $\theta_{\text{SHE}}^{\text{Pt}}$) from the reference series are found in excellent agreement with our previous studies on a different sample series \cite{Krishnia2023} and with other literature \cite{zhang2015}.  

\medskip

While the SOT fields and related parameters in Co(2)|Pt($t_{\rm{Pt}}$) and Co(2)|Pt($t_{\rm{Pt}}$)|Al*(1) series are in very good agreement with previous studies, we observe in Fig.$\ref{filtering}$(a) a completely distinct behavior when a Cu*(3) layer is deposited on top of Co(2)|Pt($t_{\rm{Pt}}$) bilayers. In order to confirm this somehow unexpected trend, we have prepared and characterized (at different time) two series with similar Co(2)|Pt($t_{\rm{Pt}}$)|Cu*(3) stacking (deposited on Si|SiO$_2$) that are shown in red circles and triangles in Fig.$\ref{filtering}$(a). For both of them, we observe a peak in the $\mu_0 H_{\rm{DL}}$/$J_{\rm{Pt}}$ evolution corresponding to a sharp increase with a maximum around $t_{\rm{Pt}}$ = 4~nm, before dropping down to the same amplitude as of the reference Co|Pt values at $t_{\rm{Pt}}$=5-6~nm. It is worth mentioning that we employed a parallel resistance model to calculate the relevant current density injected in the Pt layer estimating its resistivity $\approx$27~$\mu\Omega$~cm (see Fig\ref{RAHEandRs}(a)). This sharp enhancement as well the increase of the torque efficiency at smaller Pt thicknesses in Co(2)|Pt($t_{\rm{Pt}}$)|Cu*(3) compared to Co(2)|Pt($t_{\rm{Pt}}$) and Co(2)|Pt($t_{Pt}$)|Al*(1) cannot be understood by considering only spin Hall and/or spin interfacial Rashba effects. In consequence, this brings another unambiguous demonstration that the DL torque enhancement is due to the orbital accumulation generated at the Cu|CuOx interface and subsequently converted into a spin current by orbital-to-spin conversion through the Pt heavy metal layer.

\subsection{Contribution of the accumulation of orbital angular momentum to the magnetoresistive signal}

In order to corroborate the conclusions obtained from the torque measurements, we have also studied the evolution of the longitudinal magnetoresistance ($\Delta R_{xx}$) in the SMR geometry as a function of $t_{\rm Pt}$. In Fig.\ref{filtering}(b), we present a comparison of the SMR amplitude between Co(2)|Pt($t_{\rm Pt}$)|Cu*(3) (solid red circles) and Co(2)|Pt($t_{\rm Pt}$) (open blue squares) series in order to investigate any additional indirect orbital contribution to the conventional SMR from Co|Pt. In the case of Co|Pt, the evolution of SMR exhibits the expected non-monotonic behaviour with respect to $t_{\rm Pt}$, with an initial increase with $t_{\rm Pt}$, indicative of the increase of the spin injection efficiency from Pt, and then reaches to a peak at $t_{\rm Pt} = 3~$nm. Subsequently, the SMR decreases for $t_{\rm Pt}>3$~nm with a standard $\propto \left(\frac{1}{t_{\rm Pt}} \right)$ law due to current shunting effects for thicker Pt layer. A typical fitting function $\mathcal{F}=\eta \times \mathcal{I}$ able to reproduce such conventional SMR results is:

\begin{equation}
    \frac{\Delta R_{\rm xx}^{\rm SMR}}{R_{\rm xx}}=\mathcal{F}= \theta_{\rm SHE}^2 \left(\frac{\lambda_{sf}^{\rm Pt}}{t_{\rm Pt}}\right)\frac{\left(g_{\uparrow\downarrow}r_s\right)~\tanh^2(\frac{t_{\rm Pt}}{2\lambda_{sf}})}{1+\left(g_{\uparrow\downarrow}r_s\right)~\coth(\frac{t_{\rm Pt}}{\lambda_{sf}})},
\end{equation}

\noindent which can be then considered as a product of the spin injection efficiency at the interface $\eta$ and the so-called \textit{integral function}, $\mathcal{I}=\theta_{\rm{SHE}} \left(\frac{\lambda_{sf}^{\rm{Pt}}}{t_{\rm Pt}}\right)\tanh\left(\frac{t_{\rm{Pt}}}{2\lambda_{sf}}\right)$, describing the total current production in Pt by ISHE from the primary spin-Hall current reflected at the Pt|Co interface. The fitting to the experimental data shown as dotted lines in Fig.\ref{filtering}(b) have been performed by considering $\lambda_{sf}^{\rm Pt}=1.55\pm 0.05 $~nm, in pretty close agreement with the THz-TDS and torque data for hot carrier transport. Similar to the Cu|Al* series (see Fig.\ref{CuAl}(c)), any additional contribution from the orbital current generated at the Cu|Cu* (or Cu|CuOx) interface may give rise to an enhanced SMR signal according to $\frac{\Delta R_{\rm xx}}{R_{\rm xx}}=\frac{\Delta R_{\rm{xx}}^{\rm{SMR}}}{R_{\rm{xx}}}+\frac{\Delta R_{\rm{xx}}^{\text{add.}}}{R_{\text{add}}}$. We observe an increase of $\frac{\Delta R_{\rm{xx}}}{R_{\rm{xx}}}$  by about a factor of 2 for $t_{\rm{Pt}}=2$~nm when 3~nm Cu* deposited on top (see Fig.\ref{filtering}(b)). This MR enhancement, referred to as orbital magnetoresistance (OMR), should be understood as the effect of  orbital-to-spin conversion arising in Pt, which aligns well with the torque data. The combined sources of MR \textit{i.e.} SMR and OMR, display a typical \textit{bell} curve shape, diminishing at very small Pt thickness as well as at large Pt thickness \textit{via} the spin dissipation process. Note that the point at $t_{\rm{Pt}}=0$~nm in Fig.\ref{filtering}(b), which yields a non-zero value, corresponds to a pure OMR arising from Co(2)|Cu*(3) and its detailed mechanism should be considered different from the one previously described.

\section{Discussions and Conclusions}

Our thorough investigation of torques and magnetoresistance properties in Ta|Pt|Co|Cu|Al* and Co|Pt|Cu* series allow us to conclude that both types of oxidized interface with Cu, either Cu|Al* and Pt|Cu* possess the property to generate sizeable OAM, herafter diffusing into Pt (Co|Pt|Cu*) or into Co (Co|Cu|Al*) to produce enhanced torques. Such OAM generated can be most probably attributed to the interfacial OREE, as shown by Density Functional Theory calculations~\cite{Otani2021}. Even if bulk Cu itself may also lead to the generation of OAM through OHE, we can exclude such a predominant contribution as the thickness of the metallic Cu remains very small in all our sample series. In our samples, it became  clear that the Cu layer also serves as a transport channel for OAM from its oxidized interface towards active Co layer within the structure, as emphasized in previous study.~\cite{Otani2021}.

\medskip

In particular, one of our striking results is the demonstration of a strong enhancement of spin-orbital torques in Co|Pt|Cu* systems compared to Co|Cu*. This result questions about the inner mechanism of orbital-to-spin conversion occurring in the thin intermediate Pt layer. Even if a precise theoretical description of the experimental results is beyond the scope of the present study, our first modelling points out the particular role played by the \textit{s-d} spin-polarized electronic wave functions in transition metals strongly hybridized as well as related \textit{s-d} diffusion processes like proposed in the modelling of anisotropic magnetoresistance (AMR) effects~\cite{kokado2012}.

\medskip

In conclusion, we have demonstrated the significant role of the orbital Rashba effect to the effective current induced torques in Pt|Co|Cu|Al* and Co|Pt|Cu* (or Co|Pt|Cu|CuOx) structures, revealing previously unexplored facets of conduction electron behavior. Through a systematic investigation of thickness dependence in the Co|Pt|Cu|CuOx series, we have extracted the key physical parameters such as the efficiency of spin injection and related spin relaxation lengths as well as additional contributions that we assigned to the orbital component. We have then provided an experimental evidence for the presence of the orbital Rashba effect at the Cu|CuOx interface involving light elements, alongside SHE in Pt. Combining two added contributions allows us to achieve a twofold enhancement in the effective torques acting on the Co magnetization, highlighting the importance of including the electron's orbital degree of freedom in the material strategy to improve drastically the net amplitude of current-induced torques. Beyond the fundamental interest, this insight opens up promising avenues for controlling magnetization in spin-based devices, potentially leading to benefits in terms of energy efficiency and eco-friendly technology applications.


\section*{Acknowledgments}

We acknowledge Michel Viret and Jean-Baptiste Moussy (CEA-SPEC, Gif-sur-Yvette) for fruitful discussions and Sukhdeep Dhillon (LPENS, Paris) for his help in THz-TDS spectroscopy measurements. This study has been supported by the French National Research Agency under the project 'ORION' ANR-20-CE30-0022-02. This work contains results obtained from the experiments performed at the Ernst Ruska-Centre (ER-C) for Microscopy and Spectroscopy with Electrons at the Forschungszentrum Jülich (FZJ) in Germany. The ER-C beam-time access was provided via the DFG Core Facility Project ER-C E-014.
. 
\section*{References}

\providecommand{\noopsort}[1]{}\providecommand{\singleletter}[1]{#1}%


\newpage

\end{document}